\documentclass[superscriptaddress,twocolumn,aps,pra,10pt]{revtex4-2}
\usepackage{graphicx}
\usepackage{amsmath,amsthm,amssymb,dsfont,xcolor,soul}
\usepackage{mdframed}
\usepackage{orcidlink}
\newcommand{\ket}[1]{\left| #1 \right\rangle}
\newcommand{\bra}[1]{\left\langle #1 \right|}
\newcommand{\proj}[1]{\ket{#1}\hskip-1mm\bra{#1}}

\newcommand{\braket}[2]{\langle #1|#2 \rangle}

\usepackage{hyperref}

\begin{document}

\title{Counterfactuality, back-action, and information gain in multi-path interferometers}
\author{Jonte R. Hance\,\orcidlink{0000-0001-8587-7618}}
\email{jonte.hance@newcastle.ac.uk}
\affiliation{School of Computing, Newcastle University, 1 Science Square, Newcastle upon Tyne, NE4 5TG, UK}
\affiliation{Quantum Engineering Technology Laboratories, Department of Electrical and Electronic Engineering, University of Bristol, Woodland Road, Bristol, BS8 1US, UK}
\author{Tomonori Matsushita\,\orcidlink{0000-0001-7713-5374}}
\affiliation{Graduate School of Advanced Science and Engineering, Hiroshima University, Kagamiyama 1-3-1, Higashi Hiroshima 739-8530, Japan}
\author{Holger F. Hofmann\,\orcidlink{0000-0001-5649-9718}}
\email{hofmann@hiroshima-u.ac.jp}
\affiliation{Graduate School of Advanced Science and Engineering, Hiroshima University, Kagamiyama 1-3-1, Higashi Hiroshima 739-8530, Japan}

\begin{abstract}
The presence of an absorber in one of the paths of an interferometer changes the output statistics of that interferometer in a fundamental manner. Since the individual quantum particles detected at any of the outputs of the interferometer have not been absorbed, any non-trivial effect of the absorber on the distribution of these particles over these paths is a counterfactual effect. Here, we quantify counterfactual effects by evaluating the information about the presence or absence of the absorber obtained from the output statistics, distinguishing between classical and quantum counterfactual effects. We identify the counterfactual gain which quantifies the advantage of quantum counterfactual protocols over classical counterfactual protocols, and show that this counterfactual gain can be separated into two terms: a semi-classical term related to the amplitude blocked by the absorber, and a Kirkwood-Dirac quasiprobability assigning a joint probability to the blocked path and the output port. A negative Kirkwood-Dirac term between a path and an output port indicates that inserting the absorber into that path will have a focussing effect, increasing the probability of particles arriving at that output port, resulting in a significant enhancement of the counterfactual gain. We show that the magnitude of quantum counterfactual effects cannot be explained by a simple removal of the absorbed particles, but originates instead from a well-defined back-action effect caused by the presence of the absorber in one path, on particles in other paths.
\end{abstract}

\maketitle

\section{Introduction}

The term `counterfactual' has been applied in very different ways in quantum mechanics, and the link between these disparate uses is not immediately obvious. On the one hand, the notion of `counterfactual definiteness' has been invoked as an implicit assumption of Bell inequalities and the similar class of inequalities whose violation has been used to demonstrate quantum contextuality---in this context, the `counterfactual definiteness' refers to the hypothetical assignment of outcomes to a specific measurement, when an incompatible measurement has been carried out instead. The experimental violation of the corresponding inequality can then be seen as an indication that counterfactual definiteness does not hold when dealing with quantum phenomena \cite{hance2019counterfactualrestrictions}. 
On the other hand, counterfactual computation \cite{Mitchison2001CFComputation,Hosten2006CounterComp} allows us to see what the result of a quantum computation would be without running the computation, based on interference effects from the output of the computation. This effect is based on interaction-free measurement \cite{Renninger1960,Dicke1981Interaction,Elitzur1993Bomb,Kwiat1995IFM}, where supposedly information is gained about a quantum system through measurement without any physical interaction with the system. Other `counterfactual' effects have also been demonstrated based on interaction-free measurement: counterfactual communication \cite{Noh2009CounterfactualCrypto,Salih2013Protocol,Cao2017SalihComm,Salih2022Laws} allows us to send messages (or even quantum information \cite{Li2015SalihCounterfactual,Salih2016Qubit,Salih2021EFQubit,Salih2020DetTele,Salih_2023}) optically across a channel, without any light travelling across that channel when information is sent, and counterfactual imaging \cite{White1998IFImaging,Zhang2019GhostIFM,Hance2021CFGI} allows us to see what the optical properties of a sample would be, without sending any light into that sample. All counterfactual protocols based on interaction-free measurement make use of interference effects to work, and the role of blocked paths, on the interference effects observed in the output statistics of particles that did not take these paths, are what originally led to these protocols being referred to as `counterfactual'. However, the counterfactuality of these protocols has been heavily debated \cite{Vaidman2014SalihCommProtocol,Salih2014ReplyVaidmanComment,Griffith2016Path,Arvidsson2016Communication,Salih2018CommentPath,Griffiths2018Reply,Vaidman2019Analysis,Aharonov2019Modification,Aharonov2020ModAngMom,Wander2021Three,Salih2022Laws,Hance2021Quantum}.
Ref.~\cite{Hance2021Quantum} attempted to clarify the role of quantum effects in such counterfactual protocols, comparing two of the protocols for counterfactual quantum communication to classical methods of communicating using counterfactuals, to identify whether there was anything special that set apart the quantum protocols from closely-related classical methods. They concluded that the quantum protocols combined (wave-like) interference effects with (particle-like) discrete detection events, making use of the wave-particle duality that is a unique feature of quantum mechanics. However, the use of specific protocols made it difficult to identify the general role or meaning of quantum interference in these protocols, and provided no quantitative measure of the difference between quantum and classical counterfactuals.

In this paper, we go back to the principle underpinning all counterfactual protocols inspired by interaction-free measurement---that there is a manifest difference between the probability distributions we observe over the outcomes of an interferometer, when comparing the case when one path within the interferometer is blocked, with the case where it isn't; and that this difference is greater than the change to the distributions just caused by the loss of particles in the blocked path alone. In Section~\ref{sec:Stat}, we quantify the counterfactual effects caused by inserting an absorber into one path of a general interferometric setup by considering the statistical distance between the output probabilities with or without the absorber. This statistical distance represents the information that can be obtained regarding the presence or absence of the absorber, where the information that can be gained by simply removing the fraction of particles that is absorbed is upper bounded by the absorption probability. Our results show that quantum interference effects can increase this information significantly, allowing us to identify the `counterfactual gain' above the classical statistical distance which is associated with these non-classical effects. A non-zero counterfactual gain is only observed when there are quantum coherences between the blocked path and the paths taken by the particles detected in the output of the interferometer. Therefore, the counterfactual gain quantifies the effect of constructive and destructive interferences between the counterfactual result and its alternatives on the output distribution of photons that seemingly do not interact with the absorber. This provides an intuitive indicator of the difference between the quantum scenario and the classical notion of absorption as a removal of particles without any effects on the remaining particles. 

We separate this counterfactual gain into two terms: one term associated with the amplitude of the blocked path in the output port, referred to as Elitzur-Vaidman term due to its role in the original Elitzur-Vaidman bomb tester protocol, and one term corresponding to a Kirkwood-Dirac quasiprobability  \cite{Kirkwood1933KD,Dirac1945KD,johansen2007quantum,Halpern2018KD}, which is absent in the Elitzur-Vaidman case. This Kirkwood-Dirac term represents a generally non-positive joint probability of the blocked path and the output port, corresponding directly to the reduction in each output probability caused by the removal of particles in the blocked path. This Kirkwood-Dirac term is subtracted from the counterfactual gain, reflecting the fact that a mere removal of particles does not result in any counterfactual gain. However, a negative Kirkwood-Dirac quasiprobability will increase the counterfactual gain, and a non-zero counterfactual gain is observed whenever at least one of the outcomes has a negative Kirkwood-Dirac term. 

In Section~\ref{sec:EV}, we discuss the typical characteristics of counterfactual effects observed in the output of a multi-path interferometer. We compare two cases: one equivalent to the original Elitzur-Vaidman bomb tester, where the counterfactual gain originates from one output port having probability zero of the photon arriving in it in the case without an absorber, with a Kirkwood-Dirac term of zero; and another where the initial output probabilities are all equal, and a negative Kirkwood-Dirac term results in a counterfactual focusing effect, where the presence of the absorber results in a significant increase of the detection probability at one of the output ports. The focusing effect described here shows that the presence of the absorber redirects the paths of the photons that are not absorbed. The precise form of this redistribution is analysed in Section~\ref{sec:back}, where we derive an explicit expression for the measurement back-action of the non-detection of a particle in the path of the absorber. In Section~\ref{Sect:Three-Path}, we illustrate these effects using Hofmann's three-path interferometer \cite{hofmann2023sequential}, showing how strong and counterintuitive this this counterfactual gain can be in a scenario with a negative Kirkwood-Dirac term. Finally, we summarise our argument in Section~\ref{sec:discussion}.

The counterfactual gain we demonstrate in this paper results from the loss of coherence between states, which occurs when one of the states is lost by observation---these quantum counterfactual effects cannot be explained simply by removing the contributions to the output probabilities from the photons which are lost to the absorber. Previous discussions of quantum counterfactuals only highlighted this in specific situations, whereas we show this more generally: that the underlying effect results from initial coherences which change when one option is removed. The quantitative measure of counterfactual gain demonstrates the difference between quantum contributions and classical counterfactuals without qualitative arguments about the observation of events that would not occur when the path is not blocked. Our theory of counterfactual effects thus identifies the essential quantum effects behind a much wider range of scenarios than those considered previously.

\section{Statistical analysis of counterfactual effects}\label{sec:Stat}

In this Section, we show that quantum mechanical counterfactuals differ from classical counterfactual logic due to the loss of coherence between the states that have been excluded by the observation of a negative result, and the states which are not excluded. 
This difference can be quantified in terms of a quantum advantage in the task of discriminating between the presence of an absorber and its absence in a specific path of a single-photon interferometer by looking at which output of that interferometer, if any, the photon arrived at. Since this quantum advantage is associated with quantum mechanical counterfactuals, we will refer to it as `counterfactual gain' in the following.

Let us start by considering what we would expect from a classical (particle-like) attempt to discriminate whether or not an absorber has been inserted into the arm of a single-particle interferometer. If the only effect of an absorber inserted in a path $a$ of an interferometer was the absorption of any photons which pass through that path, the observed probabilities at the output ports would just reduce in proportion to how the absorbed photons would been distributed across the outputs if there had not been a absorber. Due to the loss of coherence between the path with the absorber and the other paths, quantum mechanics predicts much greater changes to the output probabilities than this classical constraint would allow. We should therefore quantify these additional changes to the probability distribution in order to identify the counterfactual gain between quantum and classical counterfactual effects. 

\begin{figure}
    \centering
    \includegraphics[width=\linewidth]{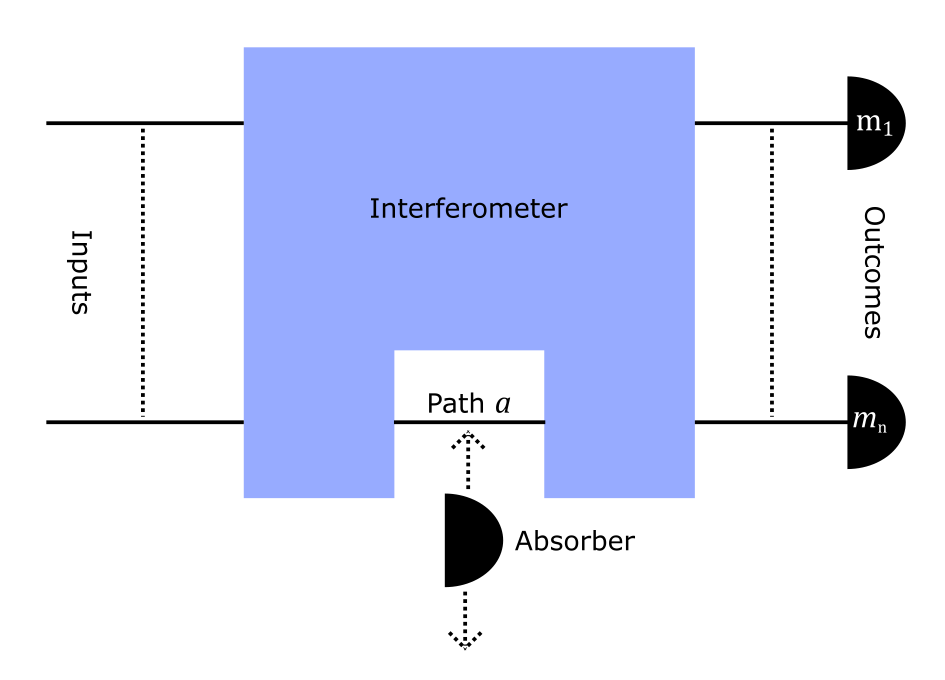}
    \caption{A generic multi-path interferometer, in a situation where path $a$ can be either blocked or not blocked by an absorber. As discussed in the text, the counterfactual scenario, where a single-photon goes through the interferometer, is not absorbed by the blocker, and goes to one of the detectors corresponding to outcomes $m$ (for $m$ between $m_1$ and $m_n$, can and should have differing probability distributions for each of the outcomes $m$ than the case where the absorber is not placed in path $a$.
    }
    \label{fig:GenericInterferometer}
\end{figure}

Without any absorber, the output probabilities for the photon arriving at each output port $m$ are given by $P(m)$. If all of the photons which pass through path $a$ are absorbed, the probability that the photon is absorbed is $P(a)$. The remaining photons, corresponding to a fraction of $1-P(a)$ of the total number of photons sent sequentially through the interferometer, will arrive at the output ports $m$ with probabilities of $P(m|X_a)$, where $X_a$ represents the counterfactual condition that the photon did not pass through path $a$. 

Consider a scenario where we are trying to guess correctly whether an absorber is present or absent. We are assuming this can be treated as being completely unbiased---based on the information we have, it is equiprobable whether or not the absorber was inserted. This assignment of equiprobability can be justified in a Bayesian sense, as our prior being complete ignorance as to whether or not the absorber was inserted. 
If our guess is based on a single trial, the optimal strategy is to compare the probability $P(m)$ without the absorber to the probability $P(m|X_a)$ in the presence of an absorber. If the absorber reduces the probability of the outcome ($P(m|X_a)<P(m)$), the best guess upon observing a photon arriving at output $m$ is that the absorber was absent. While classically the absorber can only either reduce the probability or keep it the same, to consider quantum scenarios we also need to take into account the possibility that the presence of the absorber somehow increases the probability of the outcome $m$ ($P(m|X_a)>P(m)$), in which case the best guess upon observing $m$ is that the absorber is present. These counter-intuitive outcomes are essential for a counterfactual gain in our ability to identify the presence of a detector: even though the photon was not absorbed, the outcomes $m$ with $P(m|X_a)>P(m)$ provide (non-classical) evidence for the presence of the absorber. We can now quantify the total effect of this counterfactual gain by finding the average error probability for the optimal guessing strategy described above. Since our guesses will be based on the most likely condition for the respective outcome, wrong guesses will occur whenever the outcome was caused by the less likely condition. If the detector is present, any detection in an outcome $m$ with a probability of $P(m|X_a)<P(m)$ will result in an error. If the detector is absent, any detection in an outcome $m$ with a probability of $P(m)<P(m|X_a)$ will result in an error. Assuming that the presence or absence of the detector is equally likely (as justified above), the average error probability is obtained by
\begin{equation}
    P_\text{error} = \frac{1}{2}\sum_m \text{Min}\left[P(m), P(m|X_a)\right]
\end{equation}
In the case of a classical counterfactual situation ($P(m|X_a)<P(m)$ for all $m$), the presence of the absorber will not be detected unless the photon is actually absorbed. 
This results in an average error probability of $(1-P(a))/2$ for a classical counterfactual situation, where the absorption simply removes some of the photons from the outputs $m$. For the evaluation of the counterfactual gain, it is convenient to express the error probability using a quantitative expression of the difference between the two probability distributions $\{P(m)\}$ of outcomes without the absorber and $\{P(a),P(m|X_a)\}$ of outcomes in the presence of the absorber, where $P(a)$ again is the probability of absorption. A commonly used measure of statistical distance is the total variation distance, defined as one half of the sum over the absolute values of the differences between the probabilities of each outcome. In the present case, the outcome $a$ indicates absorption and has a probability of zero in the absence of the absorber. The statistical distance is therefore given by
\begin{equation}
\label{eq:distance}
    \Delta_a \equiv \frac{1}{2} P(a) + \frac{1}{2} \sum_m \left|P(m)-P(m|X_a)\right|,
\end{equation}
so that the error probability can be given as
\begin{equation}
    P_\text{error} =  \frac{1}{2} - \frac{1}{2} \Delta_a.
\end{equation}
The statistical distance between the probabilities with and without the absorber describes the suppression of errors in the guessing game described above. In the classical counterfactual case, the statistical distance $\Delta_a$ is equal to the probability of absorption given by $P(a)$. 

The counterfactual gain, caused by the presence of quantum coherence between the absorbed path $\ket{a}$ and the other paths in the interferometer, is given by the difference between the statistical distance $\Delta_a$ and the probability of absorption $P(a)$. As explained above, it originates from outcomes $m$ with $P(m|X_A)>P(m)$. Using Eq.~(\ref{eq:distance}), we find that the counterfactual gain can be expressed as
\begin{equation}\label{eq:deltaAcfgaingen}
\Delta_a - P(a) = \sum_{P(m|X_A)>P(m)} \left( P(m|X_a) - P(m)\right).
\end{equation}
If the presence of an absorber in path $a$ increases the probability that photons in other paths arrive at output $m$, obtaining that outcome is evidence for the presence of the absorber and the increase in probability is added to the statistical distance that describes the effects of the absorber on the observable statistics. On closer inspection, the counterfactual gain corresponds to the difference between the increase in the probability of detecting the absorber gained by assigning absorber detection to the outcome $m$, and an increase in the probability of false positives associated with the probability $P(m)$ of obtaining the same result in the absence of the absorber.  

We can now apply the statistical analysis to a single particle interferometer of arbitrary size (as we show in Fig.~\ref{fig:GenericInterferometer}).
The interference effects that characterise the interferometer are given by the inner products of the Hilbert space vectors representing the paths. For any input state $\hat{\rho}$, the counterfactual output statistics are given by
\begin{equation}
\begin{split}\label{eq:cfoutputstats}
P(m|X_a) &=\bra{m} \left(\mathds{1}-\proj{a}\right) \hat{\rho} \left(\mathds{1}-\proj{a}\right) \ket{m}
\\
&= P(m) - 2 \varrho(a,m) + \left|\braket{m}{a}\right|^2 P(a),
\end{split}
\end{equation}
where
\begin{equation}
\varrho(a,m) = \mbox{Re} \left[\braket{m}{a}\hspace{-2pt}\bra{a} \hat{\rho} \ket{m} \right]
\end{equation}
is the real part of the Kirkwood-Dirac quasiprobability describing the joint quantum statistics of the two outcomes $a$ and $m$. The counterfactual gain is then given by
\begin{equation}\label{eq:Deltaa}
\begin{split}
\Delta_a - &P(a) =\\
&\sum_{P(m|X_A)>P(m)} \left(\left|\braket{m}{a}\right|^2 P(a)- 2 \varrho(a,m)\right).
\end{split}
\end{equation}
The condition for inclusion in the sum can also be given in terms of the positivity of the contribution,
\begin{equation}\label{eq:CFGcondition}
\left|\braket{m}{a}\right|^2 P(a) > 2 \varrho(a,m).
\end{equation}
The outcome $m$ contributes to the counterfactual gain whenever the Kirkwood-Dirac term $\varrho(a,m)$ is below a probability product representing the probability of a sequential non-absorbing observation of the photon, first at $a$ and then at $m$. 

In the original counterfactual scenario of Elitzur and Vaidman \cite{Elitzur1993Bomb}, the counterfactual gain was obtained for an outcome with $P(m)=0$, representing the additional condition that false positives should be avoided. In that case, the corresponding Kirkwood-Dirac term is zero as well, so that the counterfactual gain is represented entirely by the sequential probabilities $|\braket{m}{a}|^2 P(a)$. In the following, we therefore refer to these contributions to the counterfactual gain as the Elitzur-Vaidman terms.

Note that the Kirkwood-Dirac term is a quasiprobability of two non-commuting observables, which can be obtained directly through weak measurement and postselection \cite{Bamber2014Dirac}, or indirectly through strong measurement \cite{Wagner2024KDCircuit}. As a quasiprobability of $a$ and $m$, the roles of the $a$- and $m$-terms in the quasiprobability are equivalent and interchangeable. The Kirkwood-Dirac contribution has the appearance of a joint probability for the reversible sequence of events $(a,m)$, but it can take negative values. The Elitzur-Vaidman term can instead be interpreted as the probability of a regular sequence of events. It is not symmetric ($m$ definitely occurs after $a$), but is necessarily positive. We can therefore associate the Kirkwood-Dirac term with the non-classical correlations between $a$ and $m$ in the initial state, while the Elitzur-Vaidman term is more directly related to the statistics of sequential measurements. 

Any outcome $m$ with a Kirkwood-Dirac term of zero contributes its whole positive Elitzur-Vaidman term to the counterfactual gain. It is easy to construct a scenario where this is the case, by ensuring the initial probability for $m$ is zero, in which case the Kirkwood-Dirac term is zero as well. This was the scenario given by Elitzur and Vaidman \cite{Elitzur1993Bomb}, which is usually associated with counterfactual gain in quantum systems. We observe that the interpretation of this process as interaction-free would be more consistent with the subtraction of a joint probability $\rho(a,m)$ from each outcome probability $P(m)$. To recover this classical situation in quantum mechanics, $a$ needs to be an eigenstate of $\rho$, so that the Kirkwood-Dirac term and the Elitzur-Vaidman term are equal. Eq.~(\ref{eq:distance}) then reduces to a contribution from the Elitzur-Vaidman terms, which now represent a back-action free joint probability of $a$ and $m$. In the absence of coherence between $a$ and the other paths in the interferometer, this classical joint probability lowers the outcomes $P(m|x_a)$ with respect to the probabilities $P(m)$ observed in the absence of the absorber, so there will be no counterfactual gain. If the overlap $\braket{a}{m}$ of $a$ and $m$ is nonzero, quantum coherences between $a$ and states orthogonal to it are needed to achieve a probability of $P(m)=0$. Even though it may appear to be `classical,' a Kirkwood-Dirac term with a value of zero is already an indicator of quantum coherence, which is why the Elitzur-Vaidman bomb tester is a quantum effect despite involving no negative Kirkwood-Dirac terms. However, all negative Kirkwood-Dirac terms increase the counterfactual gain beyond what the Elitzur-Vaidman term contribution alone would allow.

\section{Application to a multi-path interferometer with equally distributed outcomes}\label{sec:EV}

The typical introductory example of the Elitzur-Vaidman bomb tester is based on a balanced 50:50 Mach-Zehnder interferometer, arranged so one of the detectors has a zero detection probability when the absorber is absent ($P(m_1)=0$). As shown above, we can generalise this scenario to a multi-path interferometer, characterised by the output probability $P(m=1)=0$ and a specific probability $P(a)$ for the absorption of the photon if the absorber (the `bomb') is inserted into path $a$. For simplicity, we now assume that the remaining $N-1$ output ports each have an equal probability of receiving the photon. Likewise, the reduction in the number of photons that arrive at the $N-1$ output ports when the absorber is inserted should be equally distributed.

The essence of the Elitzur-Vaidman Bomb Tester scenario is that the presence of the absorber can be detected even when no absorption happens. If absorption is a classical process, it should be impossible to get information about whether the absorber \textit{would} have absorbed the photon if it \textit{had} been in path $a$, when the arrival at any output port $m$ shows that it had \textit{not} been absorbed in path $a$. It is interesting to ask what features of quantum mechanics allow us to probe what could have happened if the present outcome had not been observed. As Ref.~\cite{Hance2021Quantum} mentions, two classical limits of optically probing the presence of an absorber using a similar setup are to consider light as completely wavelike (which we can think about as being equivalent to sending a coherent state from the source into the first beam splitter), or to consider light as completely particle-like (which we can think about as being equivalent to the photon being on a maximal mixture of all the paths). The first case agrees with our intuition about what interference effects we should see when the bomb doesn't work (all light going to output $m_1$), and when the bomb does work (light being divided equally between the two outputs)---however, in this case, some light will always be absorbed, meaning our inference that the absorber is present is obviously not interaction-free, since light will always interact with the measured object. In the second case, we retain the 50\% probability of the bomb blowing up if it works (given the input photon is put into an equal superposition of the two paths), but detections are always distributed evenly between the two outputs, meaning there is no way to distinguish these cases where the bomb would blow up, but doesn't, from cases where the bomb wouldn't blow up. Comparing this second (mixed state) case to the actual Elitzur-Vaidman bomb tester shows that coherence between the two paths the photon could have travelled on is necessary to allow us to distinguish cases where the bomb would have blown up (but didn't) from cases where it wouldn't have blown up.

We can now define a specific version of the Elitzur-Vaidman-style scenario by selecting a value of $P(a)$. Let us consider the case of $P(a)=1/3$. The initial state can then be written as
\begin{equation} \label{eq:multiinit}
    \ket{\psi}=\frac{1}{\sqrt{3}} \ket{a} + \sqrt{\frac{2}{3}}\ket{b},
\end{equation}
where $\ket{b}$ represents an arbitrary superposition of states orthogonal to $\ket{a}$. Eq.~(\ref{eq:multiinit}) can thus be used to describe arbitrarily complicated multi-path interferometers, where only the state $\ket{a}$ is associated with a specific physical path in the interferometer. Next, we define the outcome $\ket{m_1}$ so that we satisfy the condition $P(m_1)=0$ in the absence of the absorber. The output state $\ket{m_1}$ which provides the greatest counterfactual gain for this input state $\ket{\psi}$ is
\begin{equation}
    \ket{m_1}_{EV}=\sqrt{\frac{2}{3}} \ket{a} - \frac{1}{\sqrt{3}}\ket{b}.
\end{equation}
We can define an arbitrary number of additional outputs, corresponding to the total number of paths in the interferometer. However, the main result is independent of the number of paths. The counterfactual gain of the Elitzur-Vaidman scenario is given by a single Elitzur-Vaidman term,
\begin{equation} \label{eq:multi}
    \left(\Delta_a - P(a) \right)_{EV} = |_{EV}\braket{m_1}{a}|^2 P(a).
\end{equation}
For the values given above, the counterfactual gain is $2/9$. In the Elitzur-Vaidman scenario, this is equal to the probability $P(m_1|X_a)=2/9$, corresponding to the probability of detecting the presence of the absorber without any absorption. 

Now consider the additional counterfactual gain that can be obtained when the condition $P(m_1)=0$ is lifted. We can then identify two terms within Eq.~(\ref{eq:Deltaa}) which describe the actual effects of the absorber on the output statistics, the Elitzur-Vaidman term $|\braket{m}{a}|^2 P(a)$ and the Kirkwood-Dirac term $\varrho(a,m)$. Without coherence, the Kirkwood-Dirac term is always equal to the Elitzur-Vaidman term (i.e., $|\braket{m}{a}|^2 P(a)=\varrho(a,m)$). This value corresponds to the distribution of the particles lost at $a$ in $m$. Also note that, if the Kirkwood-Dirac term was a true joint (Kolmogorovian) probability, an interaction-free removal of $a$ would result in the subtraction of the Kirkwood-Dirac term (i.e., $-\varrho(a,m)$) from the original probability $P(m)$, resulting in $P(m|X_a)<P(m)$ and the absence of any counterfactual gain. There would be no Elitzur-Vaidman term, but the negative effect of the positive Kirkwood-Dirac term would be halved. 

Counterfactual gain is a consequence of the removal of destructive interferences between $a$ and paths orthogonal to $a$ in the output $m$. For $P(m)=0$, the amplitude from $a$ and the amplitude from orthogonal paths cancel exactly. When the absorber is inserted, the amplitude from $a$ is removed and the uncompensated magnitude of the amplitude from $a$ appears in the output $m$. The detection probabilities associated with this amplitude are represented by the Elitzur-Vaidman term. (Note that this is consistent with Eq.~(\ref{eq:cfoutputstats}).) Counterfactual gain with non-vanishing Kirkwood-Dirac terms represent destructive interferences that do not cancel $P(m)$ completely, because the amplitude from path $a$ and the amplitude from the other paths are not equal in $m$. Negative Kirkwood-Dirac terms are obtained when the amplitude from path $a$ is smaller than the amplitude from the other paths with which it destructively interferes, so that the removal of the amplitude from $a$ leaves a much larger amplitude behind. A negative Kirkwood-Dirac term will therefore enhance the counterfactual gain by `uncovering' an amplitude that is much larger than the amplitude from $a$ described by the Elitzur-Vaidman term.

The original Elitzur-Vaidman scenario starts with a heavily biased probability distribution due to the condition that $P(m_1)=0$ if the absorber is absent. Without this condition, it is possible to start with an equal distribution of output probabilities over all $N$ output ports. For a nine-path interferometer, the probabilities of each outcome are $P(m)=1/9$ and the outcome $m$ with a negative Kirkwood-Dirac term $\varrho(m,a)$ can be defined as
\begin{equation}
\label{eq:KDcondition}
\ket{m_1}_{KD} = \frac{1}{\sqrt{3}} \ket{a} - \sqrt{\frac{2}{3}} \ket{b}.
\end{equation}
The Kirkwood-Dirac term for this outcome is $\varrho(m_1,a)=-1/9$. Due to its negativity, the Kirkwood-Dirac term \textit{increases} the counterfactual gain by two times its absolute value,
\begin{equation} \label{eq:KDgain}
    \left(\Delta_a - P(a) \right)_{KD} = 2 |\varrho(m_1,a)|+|_{KD}\braket{m_1}{a}|^2 P(a).
\end{equation}
It should be noted that the Elitzur-Vaidman term depends on the relation between path $a$ and the output $\ket{m_1}_{KD}$ given by Eq.~(\ref{eq:KDcondition}). Here, it has a value of $1/9$. The total counterfactual gain of this scenario is equal to $1/3$, or 1.5 times the value of the Elitzur-Vaidman scenario for $P(a)=1/3$. 

Since the basic mechanism is destructive interference between the amplitude of path $a$ and the amplitude of all the other paths, it is possible to derive the optimal counterfactual gain for any given absorption probability $P(a)$ by maximising the interference effects at the output ports. 
The details of the derivation of the maximal counterfactual gain are given in Appendix~\ref{ap:deriv}, but it leads to the result
\begin{equation}\label{eq:DeltaaminusPa}
    \left(\Delta_a - P(a)\right) \leq \frac{1}{2}\left(\sqrt{(4-3P(a))P(a)}-P(a)\right).
\end{equation}
We can see by differentiating Eq.~(\ref{eq:DeltaaminusPa}) that the counterfactual gain has an overall maximal value of 1/3, and this value can only be obtained at $P(a)=1/3$. The example given above is therefore the optimal scenario for counterfactual gain. If we limit ourselves to the Elitzur-Vaidman scenario ($P(m_1)_{EV}=0$), the optimal result is obtained with $|_{EV}\braket{m_1}{a}|^2=1-P(a)$, and the upper bound of counterfactual gain is 
\begin{equation}\label{eq:DminusPforEV}
    \left(\Delta_a - P(a)\right)_{EV} \leq P(a)\left(1-P(a)\right).
\end{equation}
Perhaps not surprisingly, the Elitzur-Vaidman scenario achieves its maximal counterfactual gain of 1/4 at $P(a)=1/2$, the fully symmetric and balanced case described by a two-path interferometer. 

The reason why the Kirkwood-Dirac term was not included in the original Elitzur-Vaidman scenario is the requirement that the output $m_1$ only has non-zero probability when the absorber is in place. $P(m_1)_{EV}=0$ means that there are no false positives. For the nine path case, the initial distribution of outcomes is
\begin{equation}
\begin{split}
P(m_1)_{EV}=0,\\
P(m_i)_{EV} = 1/8,\,i\in\{2, 3,...,9\}.
\end{split}
\end{equation}
When the absorber is inserted, the probability of absorption is $P(a)=1/3$ and the remaining outcome probabilities are
\begin{equation}
\begin{split}
        P(m_1|X_a)_{EV} = 2/9,\\
        P(m_i|X_a)_{EV} = 1/18,\,i\in\{2, 3,...,9\}.
\end{split}
\end{equation}
Even though the total probability of finding the photon in outcomes $m_2$ to $m_9$ is larger than the probability of finding it in outcome $m_1$, the redistribution of photons maximises the probability in $m_1$.   

If we allow false positives, the initial distribution can be perfectly isotropic, with
\begin{equation}
P(m_i)_{KD} = 1/9,\,i\in\{1, 2,...,9\}.
\end{equation}
The insertion of the absorber now has the effect of focusing the photons in the output onto $m_1$, even though there was no indication in the initial state of the special role $m_1$ plays in the protocol. This special role is entirely defined by $m_1$'s relation with the path $a$ into which the absorber is placed. In the optimal case, the outcome probabilities in the presence of the absorber are
\begin{equation}
\begin{split}
        &P(m_1|X_a)_{KD} = 4/9,\\
        &P(m_i|X_a)_{KD} = 1/36,\,i\in\{2, 3,...,9\}.
\end{split}
\end{equation}
The presence of the absorber has increased the probability of the photon arriving at output port $m_1$ by a factor of four. The Bayesian likelihood that the absorber was present when an outcome of $m_1$ is obtained is therefore 80 \%. Incidentally, this is equal to the probability of a false negative inferred from outcomes $m_2$ to $m_9$. 

\section{Back-action}\label{sec:back}

The reason we can detect the presence of the absorber even when the particle is not absorbed is that the presence of the absorber somehow increases the probability that a particle arrives at some of the outcomes $m$. Since the absorption process itself only removes particles, this increase in probability means that the presence of the absorber causes a redistribution of particles that are not absorbed, suggesting that absorption is not the only interaction between the absorber and the particles. Interestingly, this possibility is not considered in the original Elitzur-Vaidman scenario. Instead, the claim that non-absorption is interaction-free suggests an intrinsic difference between classical statistics and quantum statistics that might explain why the interaction-free removal of particles in $a$ can somehow increase the number of particles observed at $m$ without any redistribution of the remaining particles. A natural representation of such a relation would be a quasiprobability that can assign negative joint probabilities to paths and outcomes. The Kirkwood-Dirac distribution is a suitable candidate; its potential negativity could explain counterfactual increases in the output probabilities without any back-action-related change to the initial statistics. This makes it somewhat ironic that the Elitzur-Vaidman scenario is the one where the Kirkwood-Dirac term is always necessarily zero. By itself, the Elitzur-Vaidman term does not qualify as a quasiprobability, since it fails to reproduce the outcome statistics $P(m)$ observed in the absence of absorptions when a sum over an orthogonal set of paths $a$ is performed. Moreover, its values are necessarily positive, and the {\it addition} of the term to the modified outcome probability $P(m|X_a)$ cannot be explained by the removal of particles in $a$. 

If the absorption process merely removed a fraction $P(a)$ of the input photons, the original outcome statistics $P(m)$ could be recovered by adding the contributions that originated from photons propagating through $a$. Since the probability that a photon that starts out in path $a$ will continue to $m$ is given by $|\braket{m}{a}|^2$, the total contribution of photons passing through $a$ to the outcome $m$ should be given by the Elitzur-Vaidman term. If we assume that there is no interaction between absorbers and the photons that are not absorbed, it should be possible to recover the output probability $P(m)$ by adding the outcomes for photons that passed through $a$ to the outcomes for photons that did not pass through $a$. It is possible to verify experimentally that the Elitzur-Vaidman term describes photons that passed through $a$ by blocking all of the other paths in an inversion of the original counterfactual experiment. The addition of the outcome distributions of the two experiments shows a well-defined deviation from the original distribution,
\begin{equation}
\label{eq:decoherence}
\begin{split}
P(m|X_a) + |\braket{m}{a}|^2 P(a) =  P(m) + \chi_B(m|a)
\end{split}
\end{equation}
where the modification of the original probability distribution is given by 
\begin{equation}
\chi_B(m|a) = 2 \left(|\braket{m}{a}|^2 P(a) - \varrho(a,m)\right). 
\end{equation}
This modification $\chi_B$ quantifies the unavoidable back-action effect caused by distinguishing photons that pass through $a$ from photons that do not pass through $a$. The only case where there is no back-action effect ($\chi_B=0$) is the case where
\begin{equation}
    \left|\braket{m}{a}\right|^2P(a) = \varrho(a,m).
\end{equation}
Aside from the trivial case of $\braket{m}{a}=0$, this only happens when
\begin{equation}
    \hat{\rho}\ket{a} = P(a)\ket{a}.
\end{equation}
As expected, the back-action is a consequence of quantum coherence between $\ket{a}$ and orthogonal path states in the input. The presence of the absorber necessarily removes this coherence, whether or not the photon is actually absorbed, and this loss of coherence has an irreversible effect on the output distribution of photons that are not absorbed. The argument that the counterfactual outcomes conditioned by $X_a$ are not affected by any back-action could only be maintained by arguing that the back-action in Eq.~(\ref{eq:decoherence}) originates exclusively from photons passing through $a$. However, this assumption is problematic because the data for this contribution can be taken by inserting absorbers in all of the alternative paths. The outcome statistics of photons that passed through $a$ are necessarily obtained when there is no detector in $a$, and the back-action in this path must originate from the absorbers in the other paths. The correct division of the back-action between photons in $a$ and photons that did not pass through $a$ should be completely symmetric, so that the two counterfactual scenarios are described consistently. This can be done by considering the Kirkwood-Dirac term as a valid representation of the undisturbed joint statistics of $a$ and $m$.
When absorbers are inserted into all paths except for $a$, the back-action converts the non-positive Kirkwood-Dirac term of $a$ and $m$, which represents the joint statistics of the initial quantum state, into the Elitzur-Vaidman term, which describes the sequence of projective measurements that are actually performed,
\begin{equation}
|\braket{m}{a}|^2 P(a)  =  \varrho(a,m) + \frac{\chi_B(m|a)}{2}.
\end{equation}
This effect accounts for exactly half of the total back-action, indicating that the back-action is always equally distributed between the photons that pass through $a$ and the photons which do not. The back-action of the absorber in path $a$ on photons that are not absorbed by it is therefore described by 
\begin{equation}
\label{eq:CFbackaction}
P(a|X_a) = \left(P(m)-\varrho(a,m)\right)+\frac{\chi_B(m|a)}{2},
\end{equation}
correctly reproducing the original relation in Eq.~(\ref{eq:cfoutputstats}) 
where the subtraction of two times the Kirkwood-Dirac term is split into a term representing the removal of photons in $a$ from the outcomes $m$, while the other half represents part of the back-action effect $\chi_B/2$, which ensures that no negative probabilities will be observed. It seems characteristic of quantum counterfactuals that the quasiprobability subtracted to represent the absence of the absorbed photons in the output also appears as part of the modification of the probabilities caused by the unavoidable back-action effect that provides the counterfactual gain. 

Eq.~(\ref{eq:CFbackaction}) reduces to a particularly simple form in the Elitzur-Vaidman case, where both $P(m)$ and $\varrho(a,m)$ are zero. In that case, the Elitzur-Vaidman term describes the redistribution of photons by the unavoidable back-action associated with the presence of the absorber in $a$. It may be worrying that the absorber has an effect on particles in other paths, but the only alternative would be to interpret the addition of the Elitzur-Vaidman term as the removal of a negative joint probability that just happens to resemble the (positive) probability of a sequential measurement of $a$ and $m$. As we argued above, such an interpretation would result in different joint statistics of $a$ and $m$ when the absorber is placed in a different path $a^\prime$. Moreover, the loss of coherence between $\ket{a}$ and orthogonal components in the initial quantum state makes it difficult to argue that the interaction between the absorber and the photons is limited to the photons that it actually absorbs. Instead, the description of back-action provided in the above analysis is fully consistent with alternative placements of the absorber, as well as the effects of non-absorbing/non-demolition measurements of the photon in path $a$. 

The increase of the outcome probabilities that describe the counterfactual gain in Eq.~(\ref{eq:deltaAcfgaingen}) originate from the sum of the back-action effect $\chi_B$ and a correlation effect described by the Kirkwood-Dirac quasiprobability. One peculiar aspect revealed by the analysis above is that the counterfactual gain can be calculated by ignoring all outcome probabilities that decrease when the absorber is inserted. A back-action-free counterfactual gain would only be possible if the joint probability that a photon passes through $a$ and then arrives at $m$ were negative. Ironically, the Kirkwood-Dirac term seems to describe such a situation, suggesting that part of the counterfactual gain originates from non-classical correlations that do not depend on the magnitude of the back-action. However, this negative Kirkwood-Dirac term contributes an equal amount of counterfactual gain through the back-action, with an additional back-action-related gain contributed by the Elitzur-Vaidman term. There is no counterfactual gain without back-action, and statistical contributions from negative Kirkwood-Dirac terms are always accompanied by at least as much counterfactual gain from the back-action. In other words, back-action is necessary for the observation of counterfactual gain, while a negative Kirkwood-Dirac term is sufficient for the observation of such a gain. 

To deepen our understanding of the role of statistics, it is worth looking at the case where counterfactual gain is observed in the presence of positive Kirkwood-Dirac terms. In this case, the Elitzur-Vaidman term must be at least twice the size of the positive Kirkwood-Dirac term, so that the loss of photons due to the correlation between $a$ and $m$ is more than compensated by the redistribution of photons due to back-action. In classical statistics, a joint probability would be limited by the marginal probabilities, so that one might expect that if $\varrho(a,m)\leq P(m)$, there must be a minimal rate of false positives whenever the Kirkwood-Dirac term is greater than zero. Although quantum theory also requires a minimal rate of false positives for non-zero Kirkwood-Dirac terms, the actual quantitative bound for non-zero Kirkwood-Dirac terms is
\begin{equation}
\label{eq:bound}
\left|\varrho(a,m)\right|\leq \sqrt{P(m)\;\;\left|\braket{m}{a}\right|^2P(a)}.
\end{equation}
This bound is also the geometric mean of the initial output probability $P(m)$ and the Elitzur-Vaidman term. In the absence of back-action, the Kirkwood-Dirac term is equal to the Elitzur-Vaidman term, so the probability $P(m)$ must be equal to or larger than these two terms. As shown in Eq.~(\ref{eq:CFGcondition}), counterfactual gain requires that the Kirkwood-Dirac term is less than half the size of the Elitzur-Vaidman term, so that the back-action adds more photons to the output than the statistical correlation removes. This condition is satisfied whenever the probability $P(m)$ is smaller than one quarter of the Elitzur-Vaidman term, 
\begin{equation}
P(m) < \frac{1}{4}|\braket{m}{a}|^2 P(a).      
\end{equation}
This is a sufficient condition for the observation of counterfactual gain. Whenever the rate of false positives drops below one-quarter of the Elitzur-Vaidman term, the unavoidable back-action effects associated with the possibility of absorption in path $a$ necessarily increase the probability of the outcome $m$ more than the losses of photons at the absorber reduce it.

Counterfactual gain is primarily the result of an interaction between the absorber and photons that were not absorbed. Quantum mechanics defines this interaction by introducing a new kind of relation between the back-action and non-classical correlations between the path $a$ and the outcome $m$. The Elitzur-Vaidman term represents the purest form of back-action, since it describes the situation where no photons are observed at $m$ when there is no absorber present. Effectively, the Elitzur-Vaidman term quantifies the amount of incoherent scattering that happens when the back-action effect removes the coherence between $\ket{a}$ and the orthogonal state components. Counterfactual gains much larger than the Elitzur-Vaidman term are possible when the probability $P(m)$ is larger than the Elitzur-Vaidman term, so long as the Kirkwood-Dirac term is negative. In principle, we can obtain a counterfactual gain where the back-action contribution is only slightly higher than the statistical contribution; this occurs when the probability $P(m)$ is much larger than the Elitzur-Vaidman term. However, this limit is characterised by a high rate of false positives, making it more difficult to observe the quantum advantage achieved by the counterfactual gain. 

\section{Statistics and back-action in Hofmann's Three-Path Interferometer}\label{Sect:Three-Path}

\begin{figure}
    \centering
    \includegraphics[width=\linewidth]{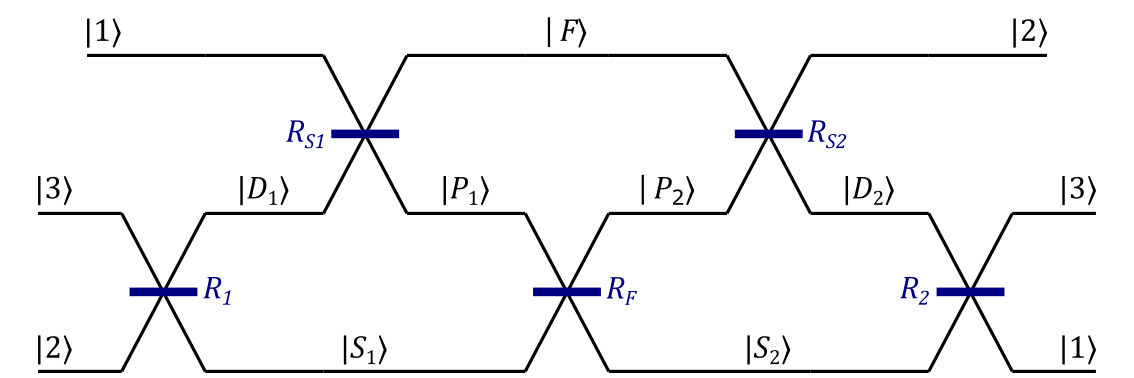}
    \caption{Hofmann's three-path interferometer \cite{hofmann2023sequential}, where paths are sequentially interfered at beamsplitters. The interferometer is aligned so that paths equivalent to the original input paths are reconstructed at the output, albeit with the position of paths $\ket{1}$ and $\ket{2}$ swapped.}
    \label{fig:threepath}
\end{figure}

It may help to illustrate the rather intricate relation between non-classical statistics and back-action effects in an interferometer specifically designed to illustrate the paradoxical aspects of quantum statistics associated with negative Kirkwood-Dirac terms \cite{hofmann2023sequential}. The layout of this interferometer, shown in Fig. ~\ref{fig:threepath}, breaks down the three-path interferences into sequential interferences at five two-path beamsplitters, identifying a total of five different measurement contexts with the different paths available to each photon. 

\begin{figure}
    \centering
    \includegraphics[width=\linewidth]{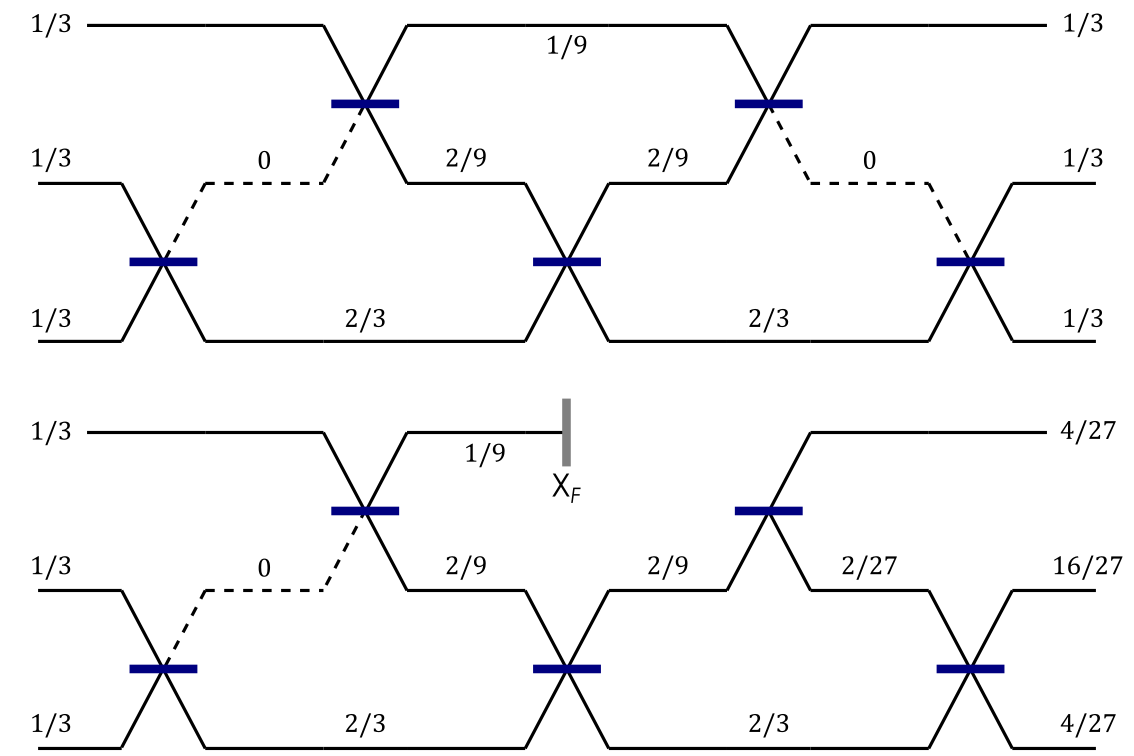}
    \caption{A comparison of the detection probabilities on different paths, in the standard three-path interferometer case, and the case where a blocker is inserted on path $\ket{F}$. These probabilities are given for the input state $\ket{N_F} = (\ket{1}+\ket{2}+\ket{3})/\sqrt{3}$.}
    \label{fig:threepathf}
\end{figure}

As we have seen above, negative Kirkwood-Dirac terms describe a specific kind of destructive interference, where the amplitude associated with the blocked path inside the interferometer is smaller than the contribution from the other paths. In the three-path interferometer, such a situation can be constructed systematically, by letting the amplitude of the path $\ket{F}$ which can be blocked by the absorber interfere destructively with the parallel path $\ket{P_2}$ in the output port $\ket{D_2}$, so that the probability of finding a photon in $\ket{D_2}$ is zero in the absence of the absorber. In the final stage of the interferometer, this empty port interferes with a third path ($\ket{S_2}$ in Fig. ~\ref{fig:threepath}), splitting into the outputs $\ket{1}$ and $\ket{3}$, while the output $\ket{2}$ is described by maximal constructive interference between $\ket{F}$ and $\ket{P_2}$. If the reflectivities of the beam splitters are laid out in the appropriate manner, the initial state can be written as an equal superposition of the outputs,
\begin{equation}
    \ket{N_F} = \frac{1}{\sqrt{3}}(\ket{1}+\ket{2}+\ket{3}),
\end{equation}
while the blocked state $\ket{F}$ is given by
\begin{equation}
    \ket{F} = \frac{1}{\sqrt{3}}(\ket{1}+\ket{2}-\ket{3}).
\end{equation}
The Kirkwood-Dirac terms of the outputs with respect to the blocked path $\ket{F}$ are 
\begin{equation}
\begin{split}
\varrho(F,1)&=1/9,
\\
\varrho(F,2)&=1/9,
\\
\varrho(F,3)&=-1/9.
\end{split}
\end{equation}
The Elitzur-Vaidman terms for the three outputs are all equal to $|\braket{m}{F}|^2 P(F)=1/27$. It may also be worth noting that the empty path $\ket{D2}$ has a Kirkwood-Dirac term of $\varrho(D_2,3)=0$ given by the sum of $\varrho(F,1)$ and $\varrho(F,3)$ and an Elitzur-Vaidman term of $|\braket{D_2}{F}|^2 P(F)=2/27$. If blocking path $F$ only removed the Kirkwood-Dirac terms, the output pattern would be $(2/9,2/9,4/9)$, with no change in $P(D_2)$. However, the back-action effect adds a redistribution of particles, so that the actual distribution of $P(m|X_F)$ is $(4/27,4/27,16/27)$ (as shown in Fig.~\ref{fig:threepathf}). Back-action reduces the lower probabilities by an additional $2/27$, and adds $4/27$ to the counterfactual gain obtained from output $\ket{3}$. The back-action also results in a non-zero particle count in $D_2$ given by $P(D_2|X_F)=2/27$. However, the total counterfactual gain of output $\ket{3}$ is equal to $7/27$, or $3.5$ times higher than the counterfactual gain of $D_2$. This massive amplification of the blocking effect serves as a useful example of the practical benefits of this work, which could have potential uses in metrology or imaging of optically-sensitive samples.

Is it possible to distinguish the effects of non-classical statistics from the additional back-action effects by considering the relation between measurement contexts? One might object that the negative value of the Kirkwood-Dirac term $\varrho(F,3)=-1/9$ could be replaced by a different quasiprobability able to explain the counterfactual effects. As shown in the previous section, this is only prevented by the observation that the Elitzur-Vaidman terms represent the propagation from $F$ to the output ports when the other ports (e.g. $P_2$ and $S_2$) are blocked, and these terms would have to be described by the same quasiprobability in order to explain the counterfactual effects. In the present scenario the Elitzur-Vaidman terms for all outputs are $1/27$, so the prediction for a removal of these particles would be a homogeneous reduction of all output probabilities to $8/27$. Conversely, the distribution of particles from $F$ in the outputs would have to be $(5/27,5/27,-7/27)$ according to the counterfactual effects observed when $F$ is blocked to explain the scenario in this way. This would be nonsensical, as path $F$ would have to somehow distribute -7/27 of a particle to output 3. Consistency between the distributions observed when path $F$ is blocked and when only $F$ is open requires that the back-action re-distributes the particles from low probability outputs to high probability outputs, with the change in probability given by $\chi_B=(-2/27,-2/27,4/27)$.

\section{Discussion}\label{sec:discussion}

This paper was motivated by the fact that the presence of an absorber in a quantum interferometer can be detected not just by the missing particles it absorbs, but also by its effects on the particles it \textit{didn't} absorb. This effect is usually identified with the appearance of photons in originally empty output ports, as in the Elitzur-Vaidman bomb tester. Here, we have shown that a much stronger effect can be obtained when the initial output probability is not zero. This additional counterfactual gain is represented by a negative Kirkwood-Dirac term that describes the quasiprobability of the particle propagating from $a$ to $m$ when path $a$ is not blocked. When path $a$ is blocked, this quasiprobability is subtracted twice from the output probability (resulting in an increase in this probability for anomalous negative values of the Kirkwood-Dirac term) and an additional term is added corresponding to the probability that a particle is first detected in $a$ and then continues onward to $m$. This additional Elitzur-Vaidman term is associated with the decoherence caused by blocking the path, and we have shown above that the addition of the difference between the Elitzur-Vaidman term and the Kirkwood-Dirac term describes the back-action effect that must occur because the absorber at $a$ distinguishes between particles in $a$ and particles not in $a$. This back-action is the reason why quantum mechanics does not allow the removal of photons from one path without changing the propagation of photons in the other paths. 

Much confusion has been caused by the assumption that non-detection of a particle requires no interaction with the particle that was not detected. Our results show that this assumption is not consistent with the theoretical description of the process. Quantum mechanics does not limit an absorber solely to interacting with the particles that it absorbs, but necessarily requires an additional interaction term that removes any coherences between the path the absorber is on, and the other paths that the particle may actually be on. It is interesting that this back-action is defined in terms of fundamental statistics, rather than any particular dynamical process associated with this decoherence. The Kirkwood-Dirac term is a consistent expression of non-classical statistics as joint probabilities for the blocked path and each output, and the Elitzur-Vaidman terms represents the incoherent propagation of particles detected in $a$. The back-action therefore amounts to a replacement of the coherence represented by the Kirkwood-Dirac term with the lack of coherence represented by the Elitzur-Vaidman term. It is interesting to note that the coherence is described in terms of the contribution of $a$ to the output $m$, which means that this coherence term will be zero if we set the condition that there are no particles in $m$ unless the absorber is inserted (no false positives). This results in the inequality in Eq.~(\ref{eq:bound}) that defines upper and lower bounds of the Kirkwood-Dirac term based on the Elitzur-Vaidman term and the rate of false positives. Our results show that the dynamics of quantum processes are fundamentally linked to non-classical statistics, making it possible to implement quantum information processing by defining the dynamics entirely in terms of the necessary back-action caused by the obstruction of selected paths, or any other method that eliminates a specific set of possibilities in a quantum system. The key to all counterfactual operations is the well-defined characteristics of the back-action dynamics necessarily induced by these obstructions in the remaining paths.

Our work also has an interpretational component: we are keen to point out that it is strange that in quantum counterfactual scenarios we intuit that there should be no change of the physical properties in unabsorbed particles. It is misleading to say that such a change is interaction-free, given we do not know what the physics of this interaction are. The result we give above is not just a technical result---it should enhance our understanding of the effects on propagation caused by the absorption of any particle. Our work should clear up the misunderstanding that the absorber should only have an effect when it absorbs a photon, and so that the effects of this absorption  can be quantified as the results of this loss alone. Quantum counterfactual effects are a typical signature of contextuality, given they are not a simple effect of `here is a value, here is how it changes due to a physical effect', but instead show that measuring in different context produces different results. This is the same as the difference between a change in a force classically, and a change in context quantum mechanically (note, Wagner et al previously linked interaction-free measurement in a two-path Mach-Zehnder interferometer to contextuality and quantum advantage \cite{wagner2022coherence}, but didn't identify the Kirkwood-Dirac negativity-related behaviour describe). The absorber serves to insert an eigencontext. A particle which doesn't experience this eigencontext is different from a particle which does. This is the counterfactual. Unlike classical counterfactuals, the effects of quantum counterfactuals are more than just the effects of the removed particles. The argument that `if the particle didn't travel via the absorber's path, it cannot know whether or not the absorber was there' is not true when the idea of considering `which path' the particle went on only comes into existence when apply this eigencontext. Particle absorption cannot be seen as complete; it is not sufficient to define the physics of the situation, and so we also need to consider whether the photon was measured in the `absorbed or not' basis. The causal connections between the coherence at the start of the interferometer and the coherence at end of the interferometer are destroyed if we measure in the `which path' basis.

Using the ideas given above, we can argue that counterfactual effects are a manifestation of contextuality that has been overlooked in the original treatment of contextuality by Kochen and Specker \cite{Kochen1968}, causing unnecessary confusion about the nature of quantum contextuality. Specifically, Kochen and Specker have placed too much emphasis on the possibility that compatible measurements may exert an influence on each other, while downplaying the possibility that contextuality originates from the well-defined relation between the outcomes of incompatible measurement contexts \cite{Marqes2014HardyLike,Ji2023Characterization,ji2024tracing,Ji2024quantitative}. Counterfactual effects illustrate this relation between incompatible measurements by introducing a minimal change of context represented by the exclusion of a single possibility. This minimal change of context has a profound impact on the output statistics because the output ports now represent a fundamentally different measurement context with respect to the input state. The undisturbed output measurement is not compatible with any information about the excluded possibility, whether counterfactual or affirmative. As we have shown above, counterfactuals describe a necessary and well-defined relation between incompatible measurements, making it possible to characterise the relation between different contexts in terms of the back-action dynamics that relates them to each other. More specifically, counterfactual effects establish a link between contextuality and quantum interferences by identifying the modification of interference effects with the change of measurement context. 

In conclusion, quantum mechanics allows us to use counterfactual effects to probe the non-classical relations between different measurement contexts. In this paper, we have provided the theoretical framework for a complete characterisation of counterfactual effects, describing both the statistical relation between the blocked path and the output ports in the absence of the observed, and the necessary back-action effect that modifies the original statistics. It is important to understand that this back-action effect is very different from the classical assumption that an absorption merely removes a particle and does nothing else. The information gained by a non-detection of the particle requires a form of dynamics that accounts for the necessary decoherence effects described by the formalism, establishing a fundamental relation between information and dynamics that distinguishes quantum mechanics from classical physics. 

\textit{Acknowledgements ---} JRH was supported by Hiroshima University's Phoenix Postdoctoral Fellowship for Research for part of the time this paper was being written. TM was supported by JST, the establishment of university fellowships towards the creation of science technology innovation, Grant Number JPMJFS2129.

\bibliographystyle{unsrturl}
\bibliography{ref.bib}

\appendix
\begin{widetext}
\section{Derivation of Maximal Counterfactual Gain}\label{ap:deriv}

The counterfactual gain that can be achieved depends on the fraction of particles that will be absorbed, as expressed by the probability $P(a)$. For a fixed value of $P(a)$, we can find the maximal counterfactual gain by considering the square roots of the probability distributions $P(m)$ and $P(m|X_a)$ as vectors, where $P(a)$ defines the squared length of the vector $(\sqrt{P(m|X_a})$ as
\begin{equation}
    \sum_m P(m|X_a) = 1 - P(a).
\end{equation}
By normalisation, the length of the vector $(\sqrt{P(m)})$ is of course one. We can then express the statistical distance using an inner product of the sum and difference vectors,
\begin{equation}
\begin{split}
    \Delta_a = \frac{1}{2} P(a) + \frac{1}{2}\sum_m \left|\sqrt{P(m|X_a)} - \sqrt{P(m)} \right|\left(\sqrt{P(m|X_a)} + \sqrt{P(m)}\right)
\end{split}
\end{equation}
The lengths of the sum and difference vectors are given by
\begin{equation}
\sum_m (\sqrt{P(m|X_a)} \pm \sqrt{P(m)})^2 = 2 - P(a) \pm 2 \sum_m \sqrt{P(m P(m|X_a)}.
\end{equation}
The maximal statistical distance is obtained when the sum and the difference vectors are parallel to each other and the inner product is equal to a product of the two lengths,
\begin{equation}
\label{eq:proto}
\Delta_a \le \frac{1}{2} P(a) + \frac{1}{2} \sqrt{\left(2 - P(a)\right)^2 - 4 \left(\sum_m \sqrt{P(m)P(m|X_a)}\right)^2}.
\end{equation}
This upper bound is not yet complete, since it still depends on the inner product of the two vectors $(\sqrt{P(m)})$ and $(\sqrt{P(m|X_a)})$. Since the inner product is subtracted, we need to find the lower bound of 
\begin{equation}
\sum_m \sqrt{P(m)P(m|X_a)} = \sum_m \left| \braket{m}{\psi} \right| \left|\bra{m}\left(\hat{I}-\proj{a}\right)\ket{\psi}\right|.
\end{equation}
Since the sum of the absolute value is larger than the absolute value of the sum,
\begin{equation}
\sum_m \left| \braket{m}{\psi} \right| \left|\bra{m}\left(\hat{I}-\proj{a}\right)\ket{\psi}\right| \ge \left|\sum_m \braket{\psi}{m}\hspace{-3pt}\bra{m}\left(\hat{I}-\proj{a}\right)\ket{\psi}\right|.
\end{equation}
The sum of the projectors $\proj{m}$ for the outcomes $m$ is equal to the identity operator $\hat{I}$,
\begin{equation}
    \sum_m \proj{m} = \hat{I}
\end{equation}
so the lower bound is given by 
\begin{equation}
\sum_m  \sqrt{P(m)P(m|X_a)} \ge 1 - P(a).
\end{equation}
By substituting this lower bound into Eq.~(\ref{eq:proto}) we obtain the bound given by Eq.~(\ref{eq:DeltaaminusPa}),
\begin{equation}
\Delta_a \le \frac{1}{2} P(a) + \frac{1}{2} \sqrt{P(a)(4-3 P(a))}.
\end{equation}
The bound of the counterfactual gain given by Eq.~(\ref{eq:DeltaaminusPa}) is obtained by subtracting $P(a)$ on both sides. It is easy to confirm that the maximal counterfactual gain overall is a gain of $1/3$ at $P(a)=1/3$. 

For comparison, the maximal counterfactual gain obtained when the there are no false positives ($P(m)=0$ for any outcome with $P(m|X_a)>P(m)$) is obtained when $|\braket{m}{a}|^2 = (1-P(a))$, so that
\begin{equation}
\Delta_a - P(a) \le P(a)(1-P(a)).
\end{equation}
As expected, the maximal gain of $1/4$ is achieved for $P(a)=1/2$. In addition to this reduction of the maximal gain, the additional condition of no false positives used in the Elitzur-Vaidman scenario makes it more difficult to observe counterfactual gain at low absorption probability. If you are aiming to maximise counterfactual gain, as defined by statistical distance, there seems to be no reason to avoid the occurrence of false positives. 

\end{widetext}

\end{document}